\documentclass[twocolumn,english]{revtex4-1}
\usepackage[T1]{fontenc}
\usepackage[latin9]{inputenc}
\usepackage{amsmath}
\usepackage{amssymb}
\usepackage{graphicx}

\makeatletter
\usepackage{siunitx}

\makeatother

\usepackage{babel}
\begin{document}
\title{Direct calculation of the ZZ-interaction rates in the multi-mode circuit-QED}
\author{Firat Solgun and Srikanth Srinivasan}
\affiliation{IBM Quantum, IBM T. J. Watson Research Center, Yorktown Heights, NY
10598 USA}
\begin{abstract}
Hamiltonians of the superconducting qubits of Transmon type involve
non-zero ZZ-interaction terms due to their finite and small anharmonicities.
These terms might lead to the unwanted accumulation of spurious phases
during the execution of the two-qubit gates. Exact calculation of
the ZZ-interaction rates requires the full diagonalization of the
circuit Hamiltonians which very quickly becomes computationally demanding
as the number of the modes in the coupler circuit increases. Here
we propose a direct analytical method for the accurate estimation
of the ZZ-interaction rates between low-anharmonicity qubits in the
dispersive limit of the multi-mode circuit-QED. We observe very good
agreement between the predictions of our method and the measurement
data collected from the multi-qubit devices. Our method being an extension
of our previous work in \citep{Z-paper} is a new addition to the
toolbox of the quantum microwave engineers as it relates the ZZ-interaction
rates directly to the entries of the impedance matrix defined between
the qubit ports.
\end{abstract}
\maketitle

\section{Introduction}

Superconducting quantum processors are one of the leading platforms
in the race to achieve fault-tolerant quantum computation. Several
important performance figures such as the qubit coherence times, gate
and measurement fidelities have been steadily improving in the last
two decades and they reached \citep{Tony-Error-Detection,Sarah-CR-99,Maika-Parity,Maika-State-Prep,QV32}
the error threshold levels as required by the quantum error-correction
protocols \citep{Kitaev-Surface-Code,Heavy-Hex}. However keeping
the same performance for the individual components while scaling the
circuits up remains a big engineering challenge \citep{npj-Jay}.

Superconducting qubits are made of Josephson junctions which are lossless
two-terminal circuit elements. Josephson junctions provide the non-linearity
needed to obtain the qubit modes by allowing the ``supercurrent''
to flow between their terminals by tunneling while introducing minimal
loss \citep{Catelani-QPs}. The most popular superconducting qubit
Transmon \citep{Transmon-paper} is obtained by shunting the Josephson
junction with a relatively large capacitor. As such the Transmon qubit
is a nonlinear oscillator with small anharmonicity to make it insensitive
to charge fluctuations that cause dephasing. Transmon qubits are often
modeled as multi-level quantum Duffing oscillators and they are designed
to interact with each other and with their environment in the circuit-QED
architecture \citep{circuit-QED,circuit-QED-Review}.

In the circuit-QED architecture qubit interactions are mediated by
the linear and passive microwave environment that the Josephson junctions
are embedded in. Qubits are typically coupled to each other and to
the control/measurement electronics by the help of microwave components
constructed out of CPW transmission lines and their bare interactions
with the internal modes of these structures are of the exchange energy
type. When the strength of these interactions is smaller than the
detuning of the qubits from the internal modes the system is said
to be operated in the dispersive regime. It was shown in \citep{Z-paper}
that in the dispersive limit multi-mode circuit-QED systems can be
described by an effective Hamiltonian of Duffing oscillators whose
interactions with each other and with the drive lines are directly
related to the entries of the impedance matrix defined between the
qubit and drive ports. 

However when one reduces the effective Hamiltonian of the circuit
to a qubit Hamiltonian by eliminating the higher levels of the Duffing
oscillators non-zero ZZ-interaction (Ising type) terms are generated
due to the finite and small anharmonicities of the qubits in addition
to the main exchange interaction terms \citep{Jay-Juelich}. These
ZZ-interaction terms might be a nuisance for some two-qubit gate schemes
such as the CR-gate \citep{Easwar-CR} which is the most popular microwave
activated gate that creates the entanglement between the qubits via
the ZX-interaction. A non-zero ZZ-term in the qubit Hamiltonian will
cause spurious phase accumulations in the presence of spectator qubits
and will lead to the loss of the gate fidelity. 

The suppression of the ZZ-term in the qubit Hamiltonians has recently
been studied actively to improve two-qubit gate fidelities. In \citep{Houck}
a coupler design is proposed that consists of two arms one of which
is frequency tunable and the coupler suppresses the ZZ-term by the
interference of the interaction paths through each arm. It is shown
that the system can be tuned to a point where the ZZ-interaction becomes
zero but the effective exchange interaction $J$ remains finite. Originally
a similar topology was used in \citep{MIT-Yan} to make exchange interaction
zero. Other approaches include the use of qubits of opposite anharmonicities
\citep{Ansari-Plourde} and of tunable qubits \citep{Aaron} to cancel
the ZZ-interaction. More recently with a circuit topology similar
to \citep{Houck} but using non-tunable elements only \citep{AK}
showed that it is possible to suppress the ZZ-term over a relatively
large band while keeping a finite $J$-coupling rate that is useful
for running the CR-gate \citep{Easwar-CR}. Although a source of cross-talk
for the CR-gate ZZ-interactions can also mediate controlled-phase
(CZ) gates \citep{Zurich-CZ,Xu-Fei-Yan,Chu-Yan,Wei-IBM}.

In this paper we develop a method for the accurate estimation of the
ZZ rates between Transmon type low anharmonicity qubits in the multi-mode
circuit-QED. The frequency dependence of the ZZ rates is captured
by the impedance entries connecting the qubit ports hence the high
computational cost of diagonalizing multi-mode Hamiltonians is avoided
making the microwave engineering \citep{IMS-paper} of multi-mode
quantum couplers streamlined. 

In Section (\ref{sec:The-Effective-Hamiltonian}) we start with the
summary of the theory \citep{Z-paper} that our method is based on.
The method is described in Section (\ref{sec:Calculation-of-The-ZZ-Rate}).
The predictions of our theory are validated with numerical simulations
on the example circuits in Section (\ref{sec:Numerical-Examples}).
In Section (\ref{sec:Experimental-Results}) we compare the experimental
data collected from multi-qubit devices to the ZZ values calculated
with our method.

\section{\label{sec:The-Effective-Hamiltonian}The Effective Hamiltonian }

We start with an overview of the results in \citep{Z-paper} as our
calculations for the estimation of the ZZ rates in the next section
will be performed in the reference frame given by the effective Hamiltonian
derived in \citep{Z-paper}. We assume that the quantum device under
study consists of Transmon qubits connected to each other and to the
readout/control lines with the help of linear and passive microwave
components such as transmission lines. For such systems the following
effective Hamiltonian is derived in \citep{Z-paper} in the dispersive
limit of the circuit-QED:

\begin{equation}
\hat{\mathcal{H}}/\hbar=\hat{H}_{Q}+\hat{H}_{\chi}+\hat{H}_{R}\label{eq:Hamiltonian-block-diagonal}
\end{equation}
$\hat{\mathcal{H}}$ is obtained by block-diagonalizing the initial
system Hamiltonian in Eq. (17) of \citep{Z-paper} by applying a Schrieffer-Wolff
transformation. $\hat{H}_{Q}$ collects the terms corresponding to
the qubit subspace:

\[
\hat{H}_{Q}=\hat{H}_{Q}^{D}+\hat{H}_{Q}^{J}+\hat{H}_{Q}^{V}
\]
where $\hat{H}_{Q}^{D}$ is the diagonal part:

\begin{equation}
\hat{H}_{Q}^{D}=\sum\limits _{i=1}^{N}\omega_{i}\hat{b}_{i}^{\dagger}\hat{b}_{i}+\frac{\delta_{i}}{2}\hat{b}_{i}^{\dagger}\hat{b}_{i}(\hat{b}_{i}^{\dagger}\hat{b}_{i}-1)\label{eq:Diagonal-Qubit-Hamiltonian}
\end{equation}
which is the quantum Hamiltonian of $N$ Duffing oscillators of frequencies
$\omega_{i}$'s, anharmonicities $\delta_{i}$'s and annihilation(creation)
operators $\hat{b}_{i}$($\hat{b}_{i}^{\dagger}$)'s; for $1\leq i\leq N$.
The anharmonicity $\delta_{i}$ of the qubit $i$ is given by \citep{Z-paper}

\begin{align}
\delta_{i} & =-E_{C}^{(i)}\left(\frac{\omega_{J_{i}}}{\omega_{i}}\right)^{2}=-\frac{E_{C}^{(i)}}{1-2E_{C}^{(i)}/\omega_{i}}\label{eq:di-Anharmonicity}
\end{align}
where $E_{C}^{(i)}$ is the charging energy of the $i$-th qubit given
by $E_{C}^{(i)}=\frac{e^{2}}{2C_{i}}$, $C_{i}$ is the total Transmon
shunting capacitance of the qubit $i$; for $1\leq i\leq N$ and $\omega_{J_{i}}=1/\sqrt{L_{J_{i}}C_{i}}$;
$L_{J_{i}}$ being the bare junction inductance corresponding to the
qubit $i$.

Exchange couplings between qubits is given by the term $\hat{H}_{Q}^{J}$

\begin{equation}
\hat{H}_{Q}^{J}=\sum\limits _{i,j}J_{ij}(\hat{b}_{i}\hat{b}_{j}^{\dagger}+\hat{b}_{i}^{\dagger}\hat{b}_{j})\label{eq:Exchange-coupling-term}
\end{equation}
where the following formula is derived in \citep{Z-paper} for the
exchange coupling $J_{ij}$ between qubit modes $i$ and $j$

\begin{equation}
J_{ij}=-\frac{1}{4}\sqrt{\frac{\omega_{i}\omega_{j}}{L_{i}L_{j}}}\mathrm{Im}\left[\frac{Z_{ij}\left(\omega_{i}\right)}{\omega_{i}}+\frac{Z_{ij}\left(\omega_{j}\right)}{\omega_{j}}\right]\label{eq:impedance-formula}
\end{equation}
where $Z_{ij}$ is the impedance entry connecting the qubit port $i$
to the qubit port $j$. Qubit ports are defined across the Josephson
junctions of the Transmon qubits: port currents are the currents flowing
through and the port voltages are the voltages developed across the
Josephson junctions. $L_{i}$ is the inductance of the $i$-th qubit
and is related to the bare junction inductance $L_{J_{i}}$ by $L_{i}=L_{J_{i}}/(1-\frac{2E_{C}^{(i)}}{\hbar\omega_{i}})$.

The term $\hat{H}_{Q}^{V}$ couples the qubits to the voltage drives
and is given by

\begin{equation}
\hat{H}_{Q}^{V}=\sum\limits _{i=1}^{N}\sum\limits _{d=1}^{N_{D}}\varepsilon_{id}(\hat{b}_{i}-\hat{b}_{i}^{\dagger})V_{d}
\end{equation}
Here $V_{d}$ is the voltage source driving the $d$-th drive line
and there are $N_{D}$ voltage sources in total. $\varepsilon_{id}$
determines the coupling rate of the $i$-th qubit to the voltage source
$V_{d}$

\begin{equation}
\varepsilon_{id}=\sqrt{\frac{\omega_{i}}{2\hbar L_{i}}}\mathrm{Im}\left[Z_{i,p(d)}(\omega_{i})\right]\frac{e^{i\theta_{d}}C_{p(d)}}{\sqrt{1+\omega_{d}^{2}Z_{0}^{2}C_{p(d)}^{2}}}
\end{equation}
where $Z_{i,p(d)}(\omega_{i})$ is the impedance entry (evaluated
at the frequency $\omega_{i}$ of the qubit $i$) connecting the qubit
port $i$ to the drive port $p(d)$ corresponding to the voltage source
$V_{d}$. $C_{p(d)}$ is the capacitance shunting the drive port $p(d)$
and $Z_{0}$ is the characteristic impedance of the drive line. $\omega_{d}$
is the frequency of the voltage source $V_{d}$ which is assumed to
be a pure sinusoidal signal for simplicity. $\theta_{d}=\frac{\pi}{2}-\arctan(\omega_{d}Z_{0}C_{p(d)})$
is the phase factor corresponding to $V_{d}$. See \citep{Z-paper}
for the details about the circuit model (the multiport Cauer network)
used to define the qubit and drive ports.

$\hat{H}_{R}$ collects the terms corresponding to the modes of the
linear passive environment that the qubits are embedded in

\begin{equation}
\hat{H}_{R}=\hat{H}_{R}^{D}+\hat{H}_{R}^{J}+\hat{H}_{R}^{V}\label{eq:internal-modes-Hamiltonian}
\end{equation}
where $\hat{H}_{R}^{D}$ is the diagonal part given by

\begin{equation}
\hat{H}_{R}^{D}=\sum\limits _{k=1}^{M}\omega_{R_{k}}\hat{a}_{k}^{\dagger}\hat{a}_{k}+\frac{\chi_{kk}^{(R)}}{2}\hat{a}_{k}^{\dagger}\hat{a}_{k}(\hat{a}_{k}^{\dagger}\hat{a}_{k}-1)\label{eq:internal-modes-diag-Hamiltonian}
\end{equation}
Here it is assumed that there are $M$ internal modes with the annihilation(creation)
operators $\hat{a}_{k}^{(\dagger)}$'s and frequencies $\omega_{R_{k}}$'s.
We note that the internal modes have acquired anharmonicities $\chi_{kk}^{(R)}$'s
generated by the junction non-linearities and described by the self-Kerr
terms in the Eq. (\ref{eq:internal-modes-diag-Hamiltonian}) above.

Similar to the case for the qubit modes the term $\hat{H}_{R}^{J}$
in Eq. (\ref{eq:internal-modes-Hamiltonian}) gives the exchange coupling
$J_{kk'}$ between the internal modes $k$ and $k'$

\begin{equation}
\hat{H}_{R}^{J}=\sum\limits _{k,k'}J_{kk'}(\hat{a}_{k}\hat{a}_{k'}^{\dagger}+\hat{a}_{k}^{\dagger}\hat{a}_{k'})
\end{equation}
and the term $\hat{H}_{R}^{V}$ couples the internal modes to the
voltage drives

\begin{equation}
\hat{H}_{R}^{V}=\sum\limits _{k=1}^{M}\sum\limits _{d=1}^{N_{D}}\varepsilon_{kd}(\hat{a}_{k}-\hat{a}_{k}^{\dagger})V_{d}
\end{equation}

$\hat{H}_{\chi}$ holds the Kerr-type coupling terms left after the
block-diagonalization. In particular the terms generating the qubit
state dependent frequency shifts $\chi_{ik}$'s in the readout resonator
frequencies are contained in $\hat{H}_{\chi}$ given by

\begin{equation}
\hat{H}_{\chi}=\sum\limits _{i=1}^{N}\sum\limits _{k=1}^{M}\chi_{ik}\hat{b}_{i}^{\dagger}\hat{b}_{i}\hat{a}_{k}^{\dagger}\hat{a}_{k}\label{eq:H-chi}
\end{equation}

\section{\label{sec:Calculation-of-The-ZZ-Rate}Calculation of The ZZ-interaction
Rate}

\begin{figure}
\begin{centering}
\includegraphics[scale=1.25]{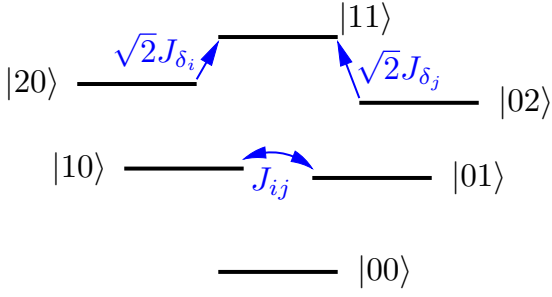}
\par\end{centering}
\caption{\label{fig:State-diagram}State diagram of the two Transmon qubits
coupled with exchange coupling as given in Eq. (\ref{eq:impedance-formula}).
The couplings $J_{\delta_{i}}$, $J_{\delta_{j}}$ of the second excited
states $\left|20\right\rangle $, $\left|02\right\rangle $ of the
qubits to the state $\left|11\right\rangle $ are updated with the
treatment of the higher-order terms. Only states with up to two qubit
excitations are shown and $\left|\Delta_{ij}\right|<\left|\delta_{i}\right|,\left|\delta_{j}\right|$
i.e. the straddling regime assumed for the detuning $\Delta_{ij}=\omega_{i}-\omega_{j}$
between the qubits $i$ and $j$.}
\end{figure}

The exchange coupling $J_{ij}$ in Eq. (\ref{eq:impedance-formula})
between qubits $i$ and $j$ can be diagonalized to get the following
qubit Hamiltonian \citep{Jay-Juelich} (with $\hbar=1$)

\begin{equation}
\hat{H}_{Q}=-\frac{(\omega_{10}+\omega_{ZZ}/2)}{2}\hat{Z}\hat{I}-\frac{(\omega_{01}+\omega_{ZZ}/2)}{2}\hat{I}\hat{Z}+\frac{\omega_{ZZ}}{4}\hat{Z}\hat{Z}\label{eq:Qubit-Hamiltonian}
\end{equation}
where $\omega_{10}=\omega_{i}+J_{ij}^{2}/\Delta_{ij}$ and $\omega_{01}=\omega_{j}-J_{ij}^{2}/\Delta_{ij}$
are the dressed qubit frequencies, $\Delta_{ij}=\omega_{i}-\omega_{j}$
the detuning between the qubits $i$ and $j$ and the ZZ-interaction
rate $\omega_{ZZ}$ is shown to be \citep{Jay-Juelich}

\begin{align}
\omega_{ZZ} & =\omega_{11}-\omega_{10}-\omega_{01}\label{eq:ZZ-definition}\\
 & =-\frac{2J_{ij}^{2}(\delta_{i}+\delta_{j})}{(\Delta_{ij}+\delta_{i})(\delta_{j}-\Delta_{ij})}\label{eq:ZZ-formula-Jay}
\end{align}
However the accuracy of this formula can be improved significantly
if one inspects more closely the higher order terms that are often
dropped with a Rotating-Wave Approximation and include in the treatment
the terms that are rotating at much slower frequencies. Such terms
will bring corrections to the couplings between the second excited
states of the qubits $\left|20\right\rangle $, $\left|02\right\rangle $
and the $\left|11\right\rangle $ state as shown in the state diagram
in Fig. (\ref{fig:State-diagram}). Note that we have $J_{\delta_{i}}=J_{\delta_{j}}=J_{ij}$
in the formula in Eq. (\ref{eq:ZZ-formula-Jay}) since the higher
order corrections are not taken into account in its derivation. To
find the higher order corrections we borrow here the fourth order
expansion of the cosine potentials of the Josephson junctions used
in Eq. (115) of \citep{Z-paper} after normal-ordering as

\begin{align}
\hat{H}_{\beta} & =\underset{pp'qq'}{-\sum}\beta_{pp'qq'}(6\hat{a}_{p}^{\dagger}\hat{a}_{p'}^{\dagger}\hat{a}_{q}\hat{a}_{q'}+4\hat{a}_{p}^{\dagger}\hat{a}_{p'}^{\dagger}\hat{a}_{q}^{\dagger}\hat{a}_{q'}+4\hat{a}_{p}^{\dagger}\hat{a}_{p'}\hat{a}_{q}\hat{a}_{q'}\nonumber \\
 & +\hat{a}_{p}\hat{a}_{p'}\hat{a}_{q}\hat{a}_{q'}+\hat{a}_{p}^{\dagger}\hat{a}_{p'}^{\dagger}\hat{a}_{q}^{\dagger}\hat{a}_{q'}^{\dagger})\label{eq:H-beta}
\end{align}
This expansion was originally given in \citep{BBQ-Yale} in a frame
different than our block-diagonal frame. Anharmonicity terms in the
diagonal Duffing Hamiltonian in Eq. (\ref{eq:Diagonal-Qubit-Hamiltonian})
and the $\chi$-shift term $\hat{H}_{\chi}$ in the Hamiltonian in
Eq. (\ref{eq:Hamiltonian-block-diagonal}) are both generated by $\hat{H}_{\beta}$
in Eq. (\ref{eq:H-beta}). To obtain the corrections to the couplings
between $\left|20\right\rangle $, $\left|02\right\rangle $ and $\left|11\right\rangle $
states as shown in Fig. (\ref{fig:State-diagram}) we need to consider
the terms $\hat{b}_{i}^{\dagger}\hat{b}_{j}^{\dagger}\hat{b}_{i}\hat{b}_{i}$
, $\hat{b}_{j}^{\dagger}\hat{b}_{i}^{\dagger}\hat{b}_{j}\hat{b}_{j}$
and their Hermitian conjugates $\hat{b}_{i}^{\dagger}\hat{b}_{i}^{\dagger}\hat{b}_{i}\hat{b}_{j}$
, $\hat{b}_{j}^{\dagger}\hat{b}_{j}^{\dagger}\hat{b}_{j}\hat{b}_{i}$;
respectively in Eq. (\ref{eq:H-beta}) above ($\hat{b}$ denotes qubit
operators in Eq. (\ref{eq:H-beta})). Note that these terms are rotating
(if one goes to the interaction picture for example) at the frequency
of qubit detuning $\Delta_{ij}$ which is much slower compared to
the other terms. The weights of these terms are $\delta_{i}\sqrt{\frac{\omega_{i}}{\omega_{j}}}\alpha_{ij}$
and $\delta_{j}\sqrt{\frac{\omega_{j}}{\omega_{i}}}\alpha_{ji}$ respectively
with

\begin{align}
\alpha_{ij} & =\frac{\mathrm{Z}_{ji}^{-1}}{2(\omega_{j}^{2}-\omega_{i}^{2})}\mathrm{Im}\left[(\omega_{i}^{2}-2\omega_{j}^{2})Z_{ij}(\omega_{j})+\omega_{i}\omega_{j}Z_{ij}(\omega_{i})\right]\label{eq:a_ij}
\end{align}
where we defined the ``cross characteristic impedance'' $\mathrm{Z}_{ij}=\sqrt{L_{i}/C_{j}}$.
The expression above is an update to the expression for $\alpha_{ij}$
originally given in Eq. (121) of \citep{Z-paper} and is calculated
using the total coordinate transformation $\boldsymbol{\alpha}=\mathbf{T}\exp(\mathbf{S})$
of \citep{Z-paper}. See Appendix (\ref{subsec:Derivation-of-the-aii-and-aij})
for the details of the derivation of the expression for $\alpha_{ij}$
in Eq. (\ref{eq:a_ij}). 

Using Eq. $\eqref{eq:a_ij}$ together with Eq. (\ref{eq:impedance-formula})
above we can now write

\begin{align}
J_{\delta_{i}} & =-\frac{1}{4}\sqrt{\frac{\omega_{i}\omega_{j}}{L_{i}L_{j}}}\mathrm{Im}\left[\alpha_{\delta_{i}}^{(i)}\frac{Z_{ij}\left(\omega_{i}\right)}{\omega_{i}}+\alpha_{\delta_{i}}^{(j)}\frac{Z_{ij}\left(\omega_{j}\right)}{\omega_{j}}\right]\label{eq:J_di}\\
J_{\delta_{j}} & =-\frac{1}{4}\sqrt{\frac{\omega_{i}\omega_{j}}{L_{i}L_{j}}}\mathrm{Im}\left[\alpha_{\delta_{j}}^{(i)}\frac{Z_{ij}\left(\omega_{i}\right)}{\omega_{i}}+\alpha_{\delta_{j}}^{(j)}\frac{Z_{ij}\left(\omega_{j}\right)}{\omega_{j}}\right]\label{eq:J_dj}
\end{align}
where the correction factors $\alpha_{\delta_{i}}^{(i)}$, $\alpha_{\delta_{i}}^{(j)}$,
$\alpha_{\delta_{j}}^{(i)}$, $\alpha_{\delta_{j}}^{(j)}$ are given
by 

\begin{align}
\alpha_{\delta_{i}}^{(i)} & =1+2\frac{\omega_{i}\delta_{i}}{(\omega_{i}^{2}-\omega_{j}^{2})}\label{eq:adii}\\
\alpha_{\delta_{i}}^{(j)} & =1-2\frac{\omega_{i}\delta_{i}}{(\omega_{i}^{2}-\omega_{j}^{2})}+4\frac{\delta_{i}}{\omega_{i}}\\
\alpha_{\delta_{j}}^{(i)} & =1+2\frac{\omega_{j}\delta_{j}}{(\omega_{i}^{2}-\omega_{j}^{2})}+4\frac{\delta_{j}}{\omega_{j}}\\
\alpha_{\delta_{j}}^{(j)} & =1-2\frac{\omega_{j}\delta_{j}}{(\omega_{i}^{2}-\omega_{j}^{2})}\label{eq:adjj}
\end{align}
Hence we can update the perturbative formula in (\ref{eq:ZZ-formula-Jay})
for the ZZ-rate as

\begin{equation}
\omega_{ZZ}=2\frac{J_{\delta_{i}}^{2}(\delta_{j}-\Delta_{ij})+J_{\delta_{j}}^{2}(\delta_{i}+\Delta_{ij})}{(\Delta_{ij}+\delta_{i})(\Delta_{ij}-\delta_{j})}\label{eq:updated-ZZ-formula}
\end{equation}
See Appendix (\ref{subsec:Derivation-of-Jdi-and-Jdj}) for the derivation
of the expressions given in Eqs. (\ref{eq:J_di}) and (\ref{eq:J_dj})
for $J_{\delta_{i}}$ and $J_{\delta_{j}}$.

There is one more term in the expansion in Eq. (\ref{eq:H-beta})
that we need to consider that contributes to the value of the ZZ-rate.
This term is of the form $\hat{b}_{i}^{\dagger}\hat{b}_{j}^{\dagger}\hat{b}_{i}\hat{b}_{j}$
which is also called as the 'cross-Kerr' term. Its contribution to
the ZZ-rate is usually small in the dispersive region. Its value can
be calculated by working out the weight coefficients $\beta_{pp'qq'}$
in (\ref{eq:H-beta}) and is given by

\begin{equation}
\omega_{ZZ}^{(K)}=2\delta_{i}(\frac{\omega_{i}}{\omega_{j}})\alpha_{ij}^{2}+2\delta_{j}(\frac{\omega_{j}}{\omega_{i}})\alpha_{ji}^{2}\label{eq:ZZ-K}
\end{equation}

Finally we introduce another parameter that was originally defined
in Eq. (120) of \citep{Z-paper}

\begin{equation}
\alpha_{ii}=\frac{1}{2}-\frac{3}{4}\mathrm{Im}[Z_{ii}(\omega_{i})]/\mathrm{Z}_{i}-\frac{1}{4}\omega_{i}\mathrm{Im}[Z'_{ii}(\omega_{i})]/\mathrm{Z}_{i}\label{eq:aii}
\end{equation}
where $\mathrm{Z}_{i}$ is the characteristic impedance of the i-th
qubit and $Z'_{ii}(\omega_{i})=\left.\frac{dZ_{ii}(\omega)}{d\omega}\right|_{\omega=\omega_{i}}$.
Note that this is an updated version of the expression given in Eq.
(120) of \citep{Z-paper}. The coefficient $\alpha_{ii}$ gives the
weight of the flux of qubit mode $i$ on its own junction's phase.
$\alpha_{ii}\cong1$ in the dispersive region but it starts to deviate
from one as the system exits the dispersive regime. Hence it is a
good measure of how dispersive the system is. To include the corrections
due to $\alpha_{ii}$'s into our treatment all we need to do is to
replace the charging energy $E_{C}^{(i)}$ in the expression for the
anharmonicity in Eq. (\ref{eq:di-Anharmonicity}) with $\alpha_{ii}^{2}E_{C}^{(i)}$;
i.e. we need to update the anharmonicity $\delta_{i}$ of the qubit
$i$ given in Eq. (\ref{eq:di-Anharmonicity}) as follows

\begin{equation}
\delta_{i}=-\frac{\alpha_{ii}^{2}E_{C}^{(i)}}{1-2\alpha_{ii}^{2}E_{C}^{(i)}/\omega_{i}}\label{eq:di-updated}
\end{equation}
The updated formula above for the anharmonicity $\delta_{i}$ allows
us to capture the frequency dependent changes due to the existence
of the high frequency modes coupled to the qubits. Refer to the Appendix
(\ref{subsec:Derivation-of-the-aii-and-aij}) for the derivation of
the expressions given in Eqs. (\ref{eq:aii})-(\ref{eq:di-updated})
for $\alpha_{ii}$ and $\delta_{i}$.

\section{\label{sec:Numerical-Examples}Numerical Examples}

We apply the method developed in the previous section for the direct
calculation of the ZZ-rates to some simple circuits and compare the
results to the exact diagonalization of the circuit Hamiltonians.

\subsection{Single Mode Coupler}

\begin{figure}
\begin{centering}
\includegraphics[scale=1.2]{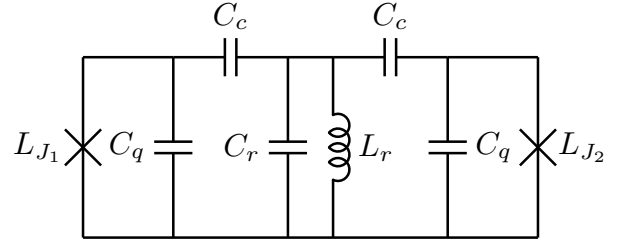}
\par\end{centering}
\caption{\label{fig:single-mode-bus}Single Mode Bus Circuit Diagram.}

\end{figure}

\begin{figure}
\begin{centering}
\includegraphics[scale=0.55]{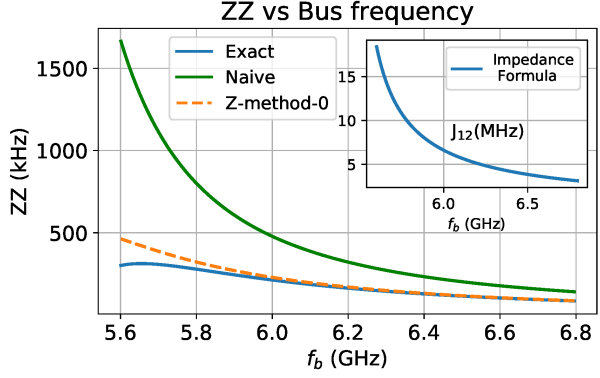}
\par\end{centering}
\caption{\label{fig:ZZ-plot-3-curves}With qubits in Fig. (\ref{fig:single-mode-bus})
fixed at $5.0\,\mathrm{GHz}$ and $5.2\,\mathrm{GHz}$ ZZ-rates calculated
using the formula in Eq. (\ref{eq:ZZ-formula-Jay}) (labeled as ``Naive''
in the figure legend) and the formula in Eq. (\ref{eq:updated-ZZ-formula})
(labeled as ``Z-method-0'' in the figure legend) are compared to
the exact value by the numerical diagonalization of the circuit Hamiltonian
(labeled as ``Exact'' in the figure legend). Inset shows the level
of $J$-coupling as a function of the bus frequency plotted applying
the impedance formula in Eq. (\ref{eq:impedance-formula}).}
\end{figure}

We start with the simple circuit shown in Fig. (\ref{fig:single-mode-bus})
consisting of two Transmon qubits coupled via a single mode shunt
$LC$ resonator bus. With $C_{q}=60\,\mathrm{fF}$,  $C_{c}=5\,\mathrm{fF}$
we fix the qubits at $5.0\,\mathrm{GHz}$ and $5.2\,\mathrm{GHz}$
by adjusting the values of $L_{J_{1}}$ and $L_{J_{2}}$ accordingly
and plot the ZZ-interaction rate as a function of the bus resonator
frequency $f_{b}$ in Fig. (\ref{fig:ZZ-plot-3-curves}) using three
different methods. The ``naive'' method of calculating the ZZ-rate
is to plug-in the value of $J_{12}$ calculated using Eq. (\ref{eq:impedance-formula})
into (\ref{eq:ZZ-formula-Jay}). However we observe in Fig. (\ref{fig:ZZ-plot-3-curves})
that there is significant discrepancy between this method (green curve
labeled ``Naive'' in the legend) and the exact value of the ZZ-rate
obtained with the numerical diagonalization of the circuit Hamiltonian
(blue curve). And this discrepancy stays at a significant level even
if we  go deep into the dispersive region; i.e. as the bus frequency
increases. However the accuracy is improved considerably when we apply
the updated formula for the ZZ-rate in (\ref{eq:updated-ZZ-formula})
by using the values of $J_{\delta_{1}}$ and $J_{\delta_{2}}$ defined
in Eqs. (\ref{eq:J_di}), (\ref{eq:J_dj}). This is plotted as ``Z-method-0''
in Fig. (\ref{fig:ZZ-plot-3-curves}) in orange color. The inset plot
informs us about the exchange coupling strength $J_{12}$ between
the qubits as a function of the bus frequency calculated using the
impedance formula in Eq. (\ref{eq:impedance-formula}) for the same
set of circuit parameters.

\begin{figure}
\begin{centering}
\includegraphics[scale=0.55]{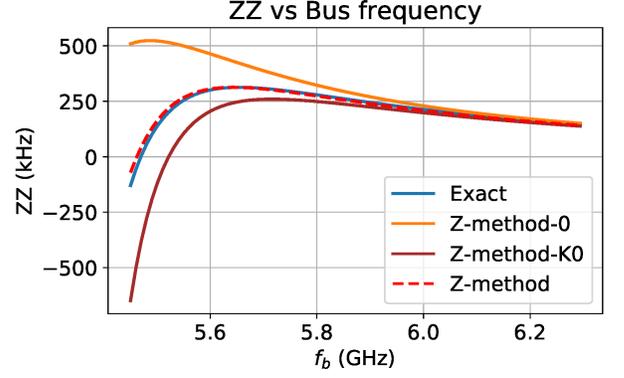}
\par\end{centering}
\caption{\label{fig:ZZ-plot-4-curves-K-aii}For the example circuit in Fig.
(\ref{fig:single-mode-bus}) with qubits fixed at 5.0 and 5.2 GHz
when only the cross-Kerr contribution in Eq. (\ref{eq:ZZ-K}) is added
(brown curved labeled ``Z-method-K0'' in the legend) to the updated
ZZ-formula in (\ref{eq:updated-ZZ-formula}) (orange) we underestimate
the exact ZZ-values (blue) slightly. However when we also include
the corrections due to the $\alpha_{ii}$ coefficients in Eqs. (\ref{eq:aii})-(\ref{eq:di-updated})
we obtain the dashed red curve(labeled as ``Z-method'' in the figure
legend) which agrees very well (even deep in the non-dispersive region)
with the exact ZZ-values (blue) obtained by a full diagonalization
of the circuit Hamiltonian.}

\end{figure}

The accuracy of our calculation of the ZZ-rate can be improved further
by adding the ``cross-Kerr'' contribution in Eq. (\ref{eq:ZZ-K})
and the corrections due to the $\alpha_{ii}$ coefficients in Eqs.
(\ref{eq:aii})-(\ref{eq:di-updated}) into our treatment. The results
are shown in Fig. (\ref{fig:ZZ-plot-4-curves-K-aii}) where we observe
that the ZZ-rate calculated with the addition of the cross-Kerm term
(brown curve) underestimates slightly the exact value (blue curve)
whereas with the addition of $\alpha_{ii}$ corrections we obtain
a very good agreement (dashed red) with true ZZ-values (blue) all
the way down to $f_{b}=5.6\,\mathrm{GHz}$ which is only $400\,\mathrm{MHz}$
away from one of the qubits.

\subsection{Two-Mode Coupler with Two $J_{12}$ Zeros}

\begin{figure}
\begin{centering}
\includegraphics[scale=0.55]{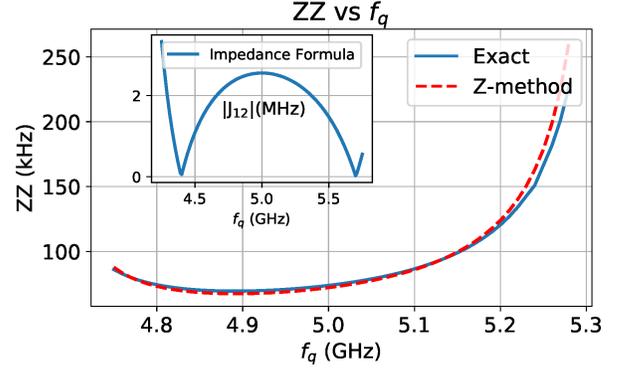}
\par\end{centering}
\caption{\label{fig:ZZ-two-mode-coupler}ZZ-rate for the trans-impedance $Z_{12}$
defined in Eq. (\ref{eq:Trans-impedance-Z12}) with one of the qubits
set to $5.0\,\mathrm{GHz}$ while the other qubit's frequency is swept
from $4.75\,\mathrm{GHz}$ to $5.30\,\mathrm{GHz}$. The exchange
coupling strength $J_{12}$ is plotted in the inset for reference.
Curves start diverging from each other after  $\sim5.2\,\mathrm{GHz}$
because the qubit detuning becomes comparable to the anharmonicity.}

\end{figure}

Here we apply our method to a coupler consisting of two finite frequency
modes. We start by artificially creating the following trans-impedance
response

\begin{equation}
Z_{12}(\omega)=\frac{A(\omega_{z_{1}}^{2}-\omega^{2})(\omega_{z_{2}}^{2}-\omega^{2})}{\omega(\omega_{p_{1}}^{2}-\omega^{2})(\omega_{p_{2}}^{2}-\omega^{2})}\label{eq:Trans-impedance-Z12}
\end{equation}
with three poles at DC, $\omega_{p_{1}}=4.0\,\mathrm{GHz}$, $\omega_{p_{2}}=6.25\,\mathrm{GHz}$
and two zeros at $\omega_{z_{1}}=4.5\,\mathrm{GHz}$ and $\omega_{z_{2}}=5.5\,\mathrm{GHz}$.
The coefficient $A$ is set to the value of $-7.97\times10^{10}.$
Assuming Transmon shunt capacitances of $65\,\mathrm{fF}$ we obtain
the results in Fig. (\ref{fig:ZZ-two-mode-coupler}).

\subsection{\label{subsec:Multi-Mode-ZZ-Cancellation-Coupler}Multi-Mode ZZ Cancellation
Coupler}

\begin{figure}
\begin{centering}
\includegraphics[scale=0.8]{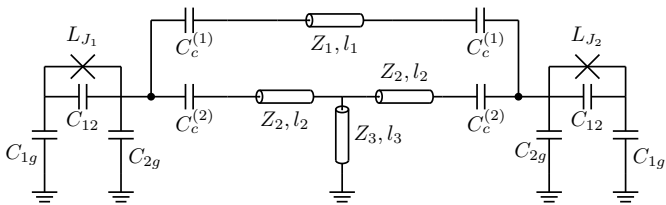}
\par\end{centering}
\caption{\label{fig:Maple-cct}Multi-mode ZZ-Cancellation Coupler Circuit.
The coupler consists of two arms: the top arm in the figure above
is a short $\lambda/2$ CPW resonator whereas the bottom arm consists
of a short CPW $\lambda/2$ section interrupted in the middle by a
$\lambda/4$ resonator shorted to ground. Both arms run in parallel
and are capacitively coupled to the same Transmon qubit pads. Circuit
parameters are $C_{12}=36\,\mathrm{fF}$, $C_{1g}=C_{2g}=46\,\mathrm{fF}$,
$C_{c}^{(1)}=8\,\mathrm{fF}$, $C_{c}^{(2)}=12\,\mathrm{fF}$. CPW
transmission lines have lengths $l_{1}=1.0\,\mathrm{mm}$, $l_{2}=0.5\,\mathrm{mm}$,
$l_{3}=3.75\,\mathrm{mm}$ and center traces of width $10\,\mathrm{um}$
with $6\,\mathrm{um}$ gap to ground which gives the characteristic
impedances $Z_{1}=Z_{2}=Z_{3}=50\,\mathrm{\varOmega}$.}

\end{figure}

\begin{figure}
\begin{centering}
\includegraphics[scale=0.55]{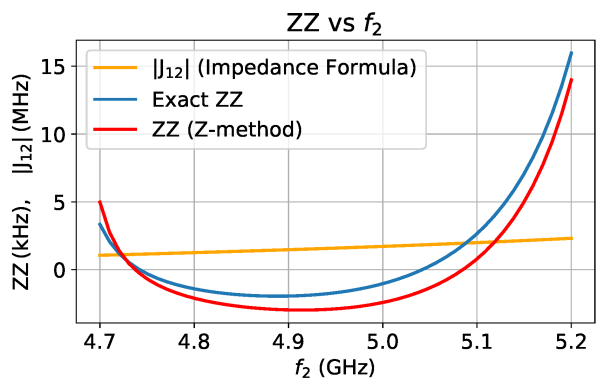}
\par\end{centering}
\caption{\label{fig:ZZ-qw-K} ZZ-interaction rate (blue and red) and the exchange
coupling rate $J_{12}$ (orange) for the coupler shown in Fig. (\ref{fig:Maple-cct})
as functions of the second qubit's frequency with the first qubit
set to $5.0\,\mathrm{GHz}$. We again compare the exact value of the
ZZ-rate to the value calculated by the impedance method applying the
formula in Eq. (\ref{eq:updated-ZZ-formula}) using the Eqs. (\ref{eq:J_di})-(\ref{eq:adjj})
and adding the cross-Kerr term in Eq. (\ref{eq:ZZ-K}) together with
the $\alpha_{ii}$ corrections (\ref{eq:aii})-(\ref{eq:di-updated}).
We observe good agreement at such a low level of ZZ-interaction and
in the location of the first zero of ZZ curves that happen at $\sim4.75\,\mathrm{GHz}$.
The second ZZ zeros are slightly off as we approach the mode at $\sim6.3\,\mathrm{GHz}$.
We included the plot of the exchange coupling $J_{12}$ which is slowly
increasing over the qubit band and reaches $2\,\mathrm{MHz}$ at $\sim5.10\,\mathrm{GHz}$.}
\end{figure}

In this section we study an example of a multi-mode coupler designed
to cancel the ZZ-interaction rate while keeping a finite $J$-coupling
strength \citep{AK}. The coupler topology consists of two branches:
one direct coupling branch of the form of a short segment of $\lambda/2$
CPW transmission line resonator and another branch of a short segment
of CPW shunted to ground in the middle through a $\lambda/4$ CPW
resonator. These two branches run in parallel and are connected to
the same Transmon qubit pads as shown in Fig. (\ref{fig:Maple-cct}).
The $\lambda/4$ section generates a mode at $\sim6.3\,\mathrm{GHz}$.
In Fig. (\ref{fig:ZZ-qw-K}) we plot ZZ-interaction rate for qubits
coupled by the circuit in Fig. (\ref{fig:Maple-cct}). ZZ-rate admits
two zeros at $\sim4.75\,\mathrm{GHz}$ and $\sim5.04\,\mathrm{GHz}$
and remains small in magnitude in between which is the typical qubit
band.

\section{\label{sec:Experimental-Results}Experimental Results}

\begin{figure}
\begin{centering}
\includegraphics[scale=0.55]{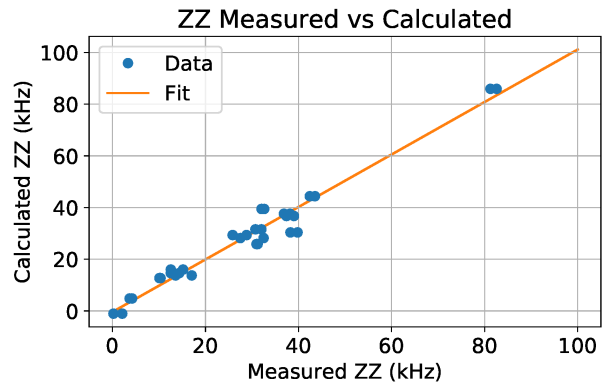}
\par\end{centering}
\caption{\label{fig:F609_C-Scatter-Plot}Scatter plot of the measured ZZ values
in the multi-qubit chip F609\_C where qubit frequencies are spread
over the band $4.8-5.0\,\mathrm{GHz}$. The least squares fit in orange
line has equation $y=1.015x-0.388$. The standard deviation is $\sigma=3.7\,\mathrm{kHz}$.
For each data point $x$-value corresponds to the measured value whereas
the $y$-value is the ZZ-rate calculated using the impedance method.
$ZZ$-rates measured from both directions are included in the plot
for each pair.}
\end{figure}

\begin{figure}
\begin{centering}
\includegraphics[scale=0.55]{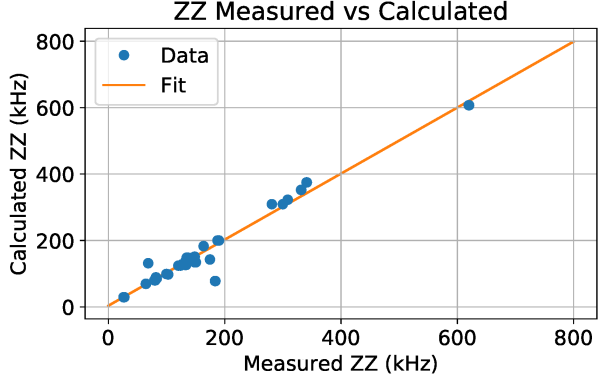}
\par\end{centering}
\caption{\label{fig:F608_C-Scatter-Plot}Scatter plot of the ZZ-values for
the Chip F608\_C. ZZ-values are larger by almost an order of magnitude
compared to the chip F609\_C in Fig. (\ref{fig:F609_C-Scatter-Plot})
since qubit frequencies are higher (in the band $5.0-5.2\,\mathrm{GHz}$).
The least squares fit line is given by $y=0.994x+3.71$. The standard
deviation is $\sigma=28.5\,\mathrm{kHz}$. Again $ZZ$-rates measured
from both directions are included in the plot for each pair when available.}
\end{figure}

In this section we test the validity of the analytical methods we
developed in the previous sections on the measurement data collected
from real devices. In Figs. (\ref{fig:F609_C-Scatter-Plot}) and (\ref{fig:F608_C-Scatter-Plot})
we compare the ZZ values predicted by our method to the measured values
from two different multi-qubit devices with qubit frequencies spread
over two different frequency bands: on chip F609\_C qubit frequencies
fall in the band $4.8-5.0\:\mathrm{GHz}$ whereas qubits on chip F608\_C
lie between $5.0-5.2\,\mathrm{GHz}$. Each chip contains 27 Transmon
qubits connected by the ZZ-cancellation type couplers introduced in
\citep{AK}. We studied this type of coupler in Section (\ref{subsec:Multi-Mode-ZZ-Cancellation-Coupler})
as an example to compare the results obtained by our analytical method
to the exact ZZ-rate values computed by the full-diagonalization of
the circuit Hamiltonian. Note that the ZZ-cancellation effect is not
observed since the cancellation band is missed in these chips. The
coupler topology stays the same across the chip in the sense that
all couplers consist of two short $\lambda/2$ CPW arms in parallel
with one of the arms shorted to ground in the middle with a $\lambda/4$
CPW section. However the length and the characteristic impedance of
the arms and the resonance frequency of the $\lambda/4$ resonator
differ creating 6 different coupler responses which allows us to validate
of our analytical methods over a larger ensemble with the good agreement
seen in Figs. (\ref{fig:F609_C-Scatter-Plot}) and (\ref{fig:F608_C-Scatter-Plot}).
The calculated ZZ-values are obtained using the $Z_{12}$ output of
the 2.5D microwave simulations evaluated at the qubit frequencies.
Self-capacitances $C_{J_{i}}$'s of the Josephson junctions are extracted
(from the anharmonicity data) to be $3.0\,\mathrm{fF}$ for the chip
F609\_C and $2.2\,\mathrm{fF}$ for the chip F608\_C as the chip F609\_C
had junctions with larger area.

\section{Conclusion}

We described a method for the accurate calculation of the ZZ-rates
between superconducting qubits in the multi-mode circuit-QED. By relating
the ZZ-interaction rates directly to the impedance entries connecting
the qubit ports our method allows streamlined analysis and design
of the qubit-qubit couplers with the help of microwave simulations.
In particular this opens a path for the design of higher-order multi-pole
ZZ-cancellation couplers by avoiding computationally intensive multi-mode
quantum Hamiltonian diagonalization.

\emph{Acknowledgements}. F. S. acknowledges support from Intelligence
Advanced Research Projects Activity (IARPA) under contract W911NF-16-1-0114-FE.

\section{Appendix}

\subsection{\label{subsec:Derivation-of-the-aii-and-aij}Derivation of the expressions
for $\alpha_{ij}$, $\alpha_{ii}$ and $\delta_{i}$}

The expressions given in Eqs. (\ref{eq:a_ij}) and (\ref{eq:aii})
for the coefficients $\alpha_{ij}$ and $\alpha_{ii}$ can be derived
by expanding the total coordinate transformation $\boldsymbol{\alpha}=\mathbf{T}\exp(\mathbf{S})$
given in \citep{Z-paper} to second-order in $\mathbf{S}$. Here $\boldsymbol{\alpha}$,
$\mathbf{T}$ and $\mathbf{S}$ are all $(N+M)\times(N+M)$ matrices
where $N$ is the number of qubits and $M$ is the number of the internal
modes. The coordinate transformation $\mathbf{T}$ is defined in Eq.
(21) of \citep{Z-paper} as 

\begin{equation}
\mathbf{T}=\left(\begin{array}{cc}
\mathbf{1}_{N\times N} & \mathbf{C}_{0}^{1/2}\mathbf{R}^{T}\\
\mathbf{0}_{M\times N} & \mathbf{1}_{M\times M}
\end{array}\right)\label{eq:T-transformation}
\end{equation}
where $\mathbf{C}_{0}$ is the diagonal matrix of Transmon capacitances
$\left(C_{1},\ldots,C_{N}\right)$  and $\mathbf{R}=\left[r_{ik}\right]$
is the $N\times M$ turns-ratio matrix of the multiport Belevitch
transformers in the canonical multiport Cauer network used in \citep{Z-paper}.
The matrix $\mathbf{S}$ defines the Schrieffer-Wolff transformation
$\exp(\mathbf{S})$ that block-diagonalizes the matrix $\mathbf{M}_{1}$
defined in Eq. (23) of \citep{Z-paper} by

\begin{eqnarray}
\mathbf{M}_{1} & = & \mathbf{T}^{T}\mathbf{C}_{0}^{-1/2}\mathbf{M}_{0}\mathbf{C}_{0}^{-1/2}\mathbf{T}\nonumber \\
 & = & \left(\begin{array}{cc}
\mathbf{\Omega}_{J}^{2} & \mathbf{\Omega}_{J}^{2}\mathbf{C}_{0}^{1/2}\mathbf{R}^{T}\\
\mathbf{R}\mathbf{C}_{0}^{1/2}\mathbf{\Omega}_{J}^{2} & \mathbf{\Omega}_{R'}^{2}
\end{array}\right)\label{eq:M1-matrix}
\end{eqnarray}
where $\mathbf{\Omega}_{J}$ is the $N\times N$ diagonal matrix holding
the qubit frequencies whereas the $M\times M$ matrix $\mathbf{\Omega}_{R'}$
corresponds to the subspace of the internal modes. $\mathbf{M}_{0}$
is the diagonal matrix with entries $\left(1/L_{1},\ldots,1/L_{N},1/L_{R_{1}},\ldots,1/L_{R_{M}}\right)$
where $L_{i}$'s are qubit inductances for $1\leq i\leq N$ and $L_{R_{k}}=1/\omega_{R_{k}}^{2}$
are the inductances of the internal modes for $1\leq k\leq M$.

After expanding $\exp(\mathbf{S})$ to second-order in $\mathbf{S}$
using for example Eqs. (B.12) and (B.15) in \citep{Winkler} we obtain
the expressions in Eqs. (\ref{eq:a_ij}) and (\ref{eq:aii}) for the
coefficients $\alpha_{ij}$ and $\alpha_{ii}$ from the $(i,j)$ and
$(i,i)$ entries, respectively, of the upper-left $N\times N$ subsector
of $\mathbf{T}\exp(\mathbf{S})$ corresponding to the qubit subspace.

The updated expression in Eq. (\ref{eq:di-updated}) for the anharmonicity
$\delta_{i}$ can be derived by noting that the capacitance re-scaling
performed by the diagonal matrix $\mathbf{C}_{0}^{-1/2}$ can be lumped
into the total coordinate transformation $\boldsymbol{\alpha}$ which
would re-normalize the diagonal capacitance $C_{i}$ by $1/\alpha_{ii}^{2}$
hence the charging energy $E_{C}^{(i)}$ gets updated by the prefactor
$\alpha_{ii}^{2}$; i.e. $E_{C}^{(i)}\rightarrow\alpha_{ii}^{2}E_{C}^{(i)}$.
One then obtains $\delta_{i}=-\frac{\alpha_{ii}^{2}E_{C}^{(i)}}{1-2\alpha_{ii}^{2}E_{C}^{(i)}/\omega_{i}}$
for the anharmonicity using the updated charging energy in the formula
given in Eq. (\ref{eq:di-Anharmonicity}).

\subsection{\label{subsec:Derivation-of-Jdi-and-Jdj}Derivation of the expressions
for $J_{\delta_{i}}$ and $J_{\delta_{j}}$}

The expression given in Eq. (\ref{eq:J_di}) for $J_{\delta_{i}}$
is obtained by noting that the term $\hat{b}_{i}^{\dagger}\hat{b}_{j}^{\dagger}\hat{b}_{i}\hat{b}_{i}$
and its Hermitian conjugate $\hat{b}_{i}^{\dagger}\hat{b}_{i}^{\dagger}\hat{b}_{i}\hat{b}_{j}$
which have weight $\delta_{i}\sqrt{\frac{\omega_{i}}{\omega_{j}}}\alpha_{ij}$
in the expansion $\hat{H}_{\beta}$ given in Eq. (\ref{eq:H-beta})
($\hat{b}$ denotes qubit operators in Eq. (\ref{eq:H-beta}) ) couple
the state $\left|20\right\rangle $ to the state $\left|11\right\rangle $
with strength $\sqrt{2}\delta_{i}\sqrt{\frac{\omega_{i}}{\omega_{j}}}\alpha_{ij}$.
This brings an additive correction to the main coupling $\sqrt{2}J_{ij}$
between the states $\left|20\right\rangle $ and $\left|11\right\rangle $
generated by the exchange coupling $J_{ij}$ given by the expression
in Eq. (\ref{eq:impedance-formula}) between qubits $i$ and $j$;
i.e. $J_{\delta_{i}}=J_{ij}+\delta_{i}\sqrt{\frac{\omega_{i}}{\omega_{j}}}\alpha_{ij}$.
After re-arrangement we arrive at the expression given in Eq. (\ref{eq:J_di})
for $J_{\delta_{i}}$. A similar analysis applied to the terms $\hat{b}_{j}^{\dagger}\hat{b}_{i}^{\dagger}\hat{b}_{j}\hat{b}_{j}$
and $\hat{b}_{j}^{\dagger}\hat{b}_{j}^{\dagger}\hat{b}_{j}\hat{b}_{i}$
(with weight $\delta_{j}\sqrt{\frac{\omega_{j}}{\omega_{i}}}\alpha_{ji}$)
for the coupling between the states $\left|02\right\rangle $ and
$\left|11\right\rangle $ gives the expression in Eq. (\ref{eq:J_dj})
for $J_{\delta_{j}}$.

\end{document}